\documentclass[conference]{IEEEtran}
\IEEEoverridecommandlockouts
\usepackage{cite}
\usepackage{amsmath,amssymb,amsfonts}
\usepackage{algorithmic}
\usepackage{graphicx}
\usepackage{textcomp}
\usepackage{mathtools}
\usepackage{amsfonts} 
\usepackage{color,soul}

\usepackage{subfigure}
\usepackage{bm}
\usepackage{float}

\def\BibTeX{{\rm B\kern-.05em{\sc i\kern-.025em b}\kern-.08em
    T\kern-.1667em\lower.7ex\hbox{E}\kern-.125emX}}
    
\begin{document}


\title{Masked Self-Supervision for Remaining Useful Lifetime Prediction in Machine Tools}

\author{
\IEEEauthorblockN{Haoren Guo}
\IEEEauthorblockA{\textit{Electrical and Computer Engineering}\\
\textit{National University of Singapore}\\
Singapore, Singapore\\
haorenguo\_06@u.nus.edu}
\\
\IEEEauthorblockN{Vadakkepat, Prahlad}
\IEEEauthorblockA{\textit{Electrical and Computer Engineering}\\
\textit{National University of Singapore}\\
Singapore, Singapore\\
prahlad@nus.edu.sg}
\and
\IEEEauthorblockN{Haiyue Zhu}
\IEEEauthorblockA{\textit{Singapore Institute of Manufacturing}\\
\textit{Technology (SIMTech), A*STAR}\\
Singapore, Singapore\\
zhu\_haiyue@simtech.a-star.edu.sg}
\\
\IEEEauthorblockN{Weng Khuen Ho}
\IEEEauthorblockA{\textit{Electrical and Computer Engineering}\\
\textit{National University of Singapore}\\
Singapore, Singapore\\
wk.ho@nus.edu.sg}
\and
\IEEEauthorblockN{Jiahui Wang}
\IEEEauthorblockA{\textit{Electrical and Computer Engineering}\\
\textit{National University of Singapore}\\
Singapore, Singapore\\
wjiahui@u.nus.edu}
\\
\IEEEauthorblockN{Tong Heng Lee}
\IEEEauthorblockA{\textit{Electrical and Computer Engineering}\\
\textit{National University of Singapore}\\
Singapore, Singapore\\
eleleeth@nus.edu.sg}
}

\maketitle

\begin{abstract}
Prediction of Remaining Useful Lifetime (RUL) 
in the modern manufacturing and automation workplace
for 
machines and tools is essential in Industry 4.0.
This is clearly evident 
as continuous tool wear, or worse, sudden machine breakdown, 
will lead to various manufacturing failures 
which would clearly cause economic loss. 
With the availability of deep learning approaches, 
the great potential and prospect of utilizing these 
for RUL prediction have resulted in
several models which are designed (for RUL prediction) 
driven by operation data of manufacturing machines. 
Current efforts in these which are based on
fully-supervised models heavily rely on the data labeled with their RULs.
However, in these cases, 
the required RUL prediction data ({\it i.e.} the annotated and labeled data from faulty and/or degraded machines) 
can only be obtained after the machine break-down occurs. 
The scarcity of broken machines in the modern manufacturing and automation workplace
in real-world situations
increases the difficulty of getting such sufficient annotated and labeled data. 
In contrast, 
the
data from healthy machines (and which are currently in operation) is much easier to be collected. 
Noting this challenge and the potential for improved effectiveness and applicability,
we thus propose (and also fully develop) a method based on the idea of masked autoencoders
which will utilize unlabeled data to do self-supervision. 
In thus the work here,
a noteworthy
masked self-supervised learning approach 
is developed and utilized;
and this is designed
to seek to build a deep learning model for RUL prediction by utilizing unlabeled data. 
The experiments to verify the effectiveness of this development
are implemented on the C-MAPSS datasets (which is collected from 
the data from the
NASA turbofan engine). 
The results rather clearly show that our development and approach here performs better, 
in both accuracy and effectiveness,
for RUL prediction 
when compared with approaches utilizing
a fully-supervised model.
\end{abstract}

\begin{IEEEkeywords}
Deep Learning, Predictive Maintenance, Remaining Useful Lifetime, Self-Supervised Learning, Time Series
\end{IEEEkeywords}

\thanks{This work has been submitted to the IEEE for possible publication.
Copyright may be transferred without notice, after which this version may no longer be accessible.}

\begin{figure}
\centering
\centerline{\includegraphics[scale = 0.1]{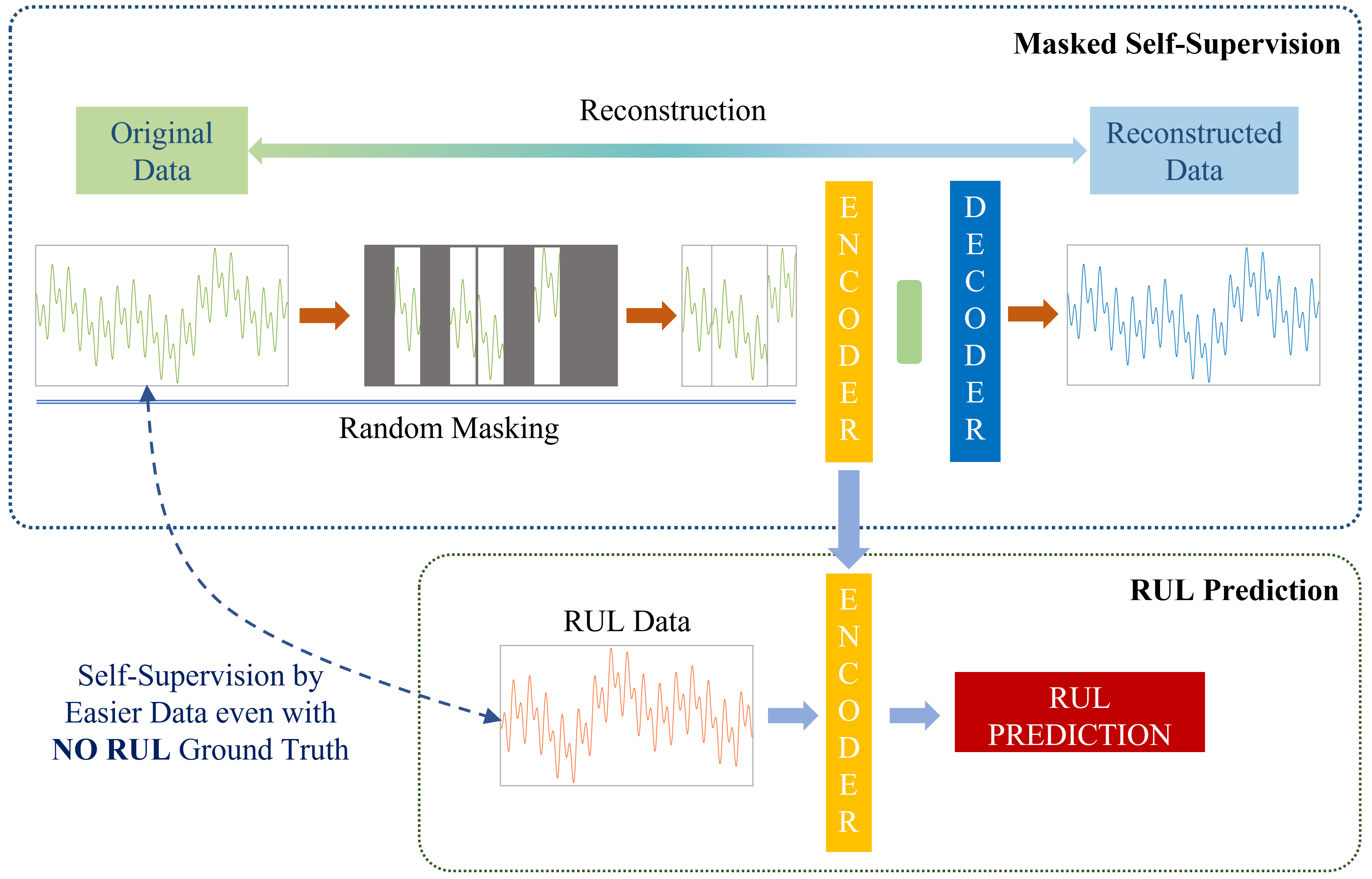}}
\caption{Framework of Masked Self-Supervision for RUL Prediction: There are two tasks in our proposed method. The upper block is the masked self-supervision task which is aiming to reconstruct the masked input to do pretraining. The lower block is the RUL prediction task which would inherit the encoder learnt from masked self-supervision to have a better feature extractor.}

\label{fig1:Architecture of Masked Self-Supervision for RUL Prediction}
\end{figure}
\section{Introduction}
Industry 4.0 has given the important impetus towards
a significant revolution in technology 
which brings the automation industry 
to evolve into a newer generation with many new technological advancements and improvements \cite{industry_4}. 
Predictive Maintenance (PDM) is an essential branch of Industry 4.0 
which could enhance the overall maintenance and reliability of the machining process 
and improve the cost efficiency \cite{PDM_survey}. 
The key points of PDM are to estimate remaining useful lifetime of equipments, 
increase safety of the industrial plant, 
decrease accidents with a negative impact on the environment, 
and do optimization on spare parts handling and workflow \cite{PDM_sub}. 
Remaining Useful Lifetime(RUL) prediction, which is to measure lifetime 
from current timestamp 
to the end of usage time of a machine or machine tools \cite{RUL_def}, 
is an important research and development topic of PDM.

In earlier times, 
the approaches adopted for RUL prediction mainly depend on physical models, 
such as acoustic, infrared, and vibration analysis\cite{PDM_physics}. 
Nowadays, with the increasing number and availablity of sensors, more data can be collected from machines. 
Therefore, along this line, there is certainly great interest in investigating and developing
an approach utilizing a data-driven model for RUL prediction
(with the very real possibility of further improved and successful development).
Thus then too, the experience-based physical models are less needed. 
Currently, many data-driven methods have been applied to do such prediction on various physical models (including machines);
and these include approaches  
such as convolutional neural network (CNN)\cite{cnn}, 
deep convolutional neural network (DCNN)\cite{dcnn2} 
and deep belief network (DBN)\cite{dbn}. 
In all these efforts,
they have achieved significant improvements on accuracy and efficiency 
compared with approaches utilizing only physical models.


\begin{figure*}
\centerline{\includegraphics[width = 14cm]{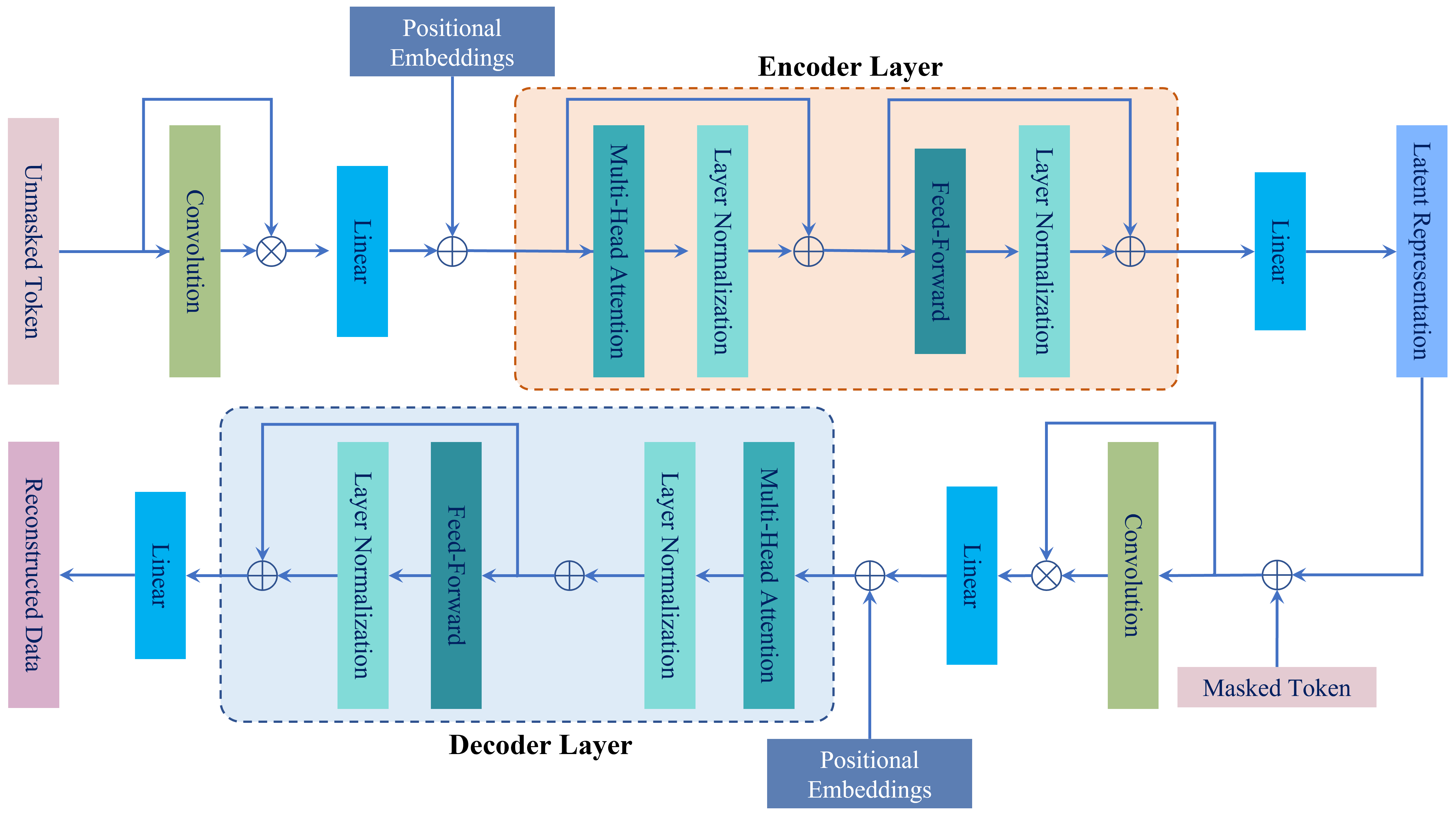}}
\caption{Architecture of Masked Self-Supervision: the unmasked tokens are encoded by a transformer after getting the semantic representation from the gated convolutional unit. The output of the encoder is the latent representation which would be restored with masked tokens and would be decoded. Similar to encoding, before decoding, the input of the decoder should be passed through a gated convolutional unit. Then, the head of the decoder will do reconstruction.}
\label{fig:Encoder and Decoder}
\end{figure*}

However, most of the current deep learning methods are approaches developed utilizing fully-supervised methodologies.  
Furthermore, current fully-supervised models heavily depend on the annotated data;
but it can be noted that it is a rather difficult constraint for these 
required annotations
(when these
are the data characterizing the remaining useful lifetime)
as these
can only be calculated backwards after the machine breakdown. 
The scarcity of broken machines in the modern manufacturing and automation workplace
in real-world situations
increases the difficulty of getting such sufficient annotated and labeled data. 
It is certainly a difficult challenge for a deep learning model to be trained well without enough data, 
and in these situations,
it is also known to be the case that
the model may get overfitted easily\cite{overfit} and the features extracted from the model may be of the under-representation type\cite{grasp}. 
With all these considerations in mind,
it is pertinent thus to note that
compared with the data from faulty and degraded machines, 
the
data from healthy machines (and which are currently in operation) is much easier to be collected. 
Also,
besides the models we mentioned above 
(which are developed for RUL predictions in earlier efforts), 
most of the later efforts are still developed 
utilizing
model-based methodologies; 
such as the added gated convolutional unit to transformer\cite{Transformer-gated-conv} approach,
and the stacked deep neural network\cite{stack_dcnn} approach. 
Along this line then,
this essentially can likely pose possible constraints and
also a likely bottleneck in RUL prediction, 
if the available efforts depend substantially 
on the difficult requirement for
approriately accurate model development.
In particular thus, 
these efforts will be highly constrained 
when encountered with the situation of
the problem of data insufficiency\cite{incre-few} and data utilization efficiency\cite{few_shot}.
In the work here, 
a proposed masked self-supervision model 
is utilized and developed to
yield a notable approach to address
these problems.

In RUL prediction, 
in current efforts and advances,
only rather relatively few works 
are developed utilizing the methodology of
self-supervised learning. 
Thus for example,
in \cite{semi-super-rul}, the work there introduced a semi-self-supervision method to predict 
the RUL differences between two timestamps. 
Interestingly too, based on the available open literature,
there appears to be
no appreciable developments designed and attempting 
to reconstruct RUL datasets from incomplete or masked data. 
Noting this challenge and the potential for improved effectiveness and applicability,
in the work here,
we thus propose (and also fully develop) a method based on the idea of masked autoencoders\cite{MAE} 
which will utilize unlabeled data to do self-supervision. 
The self-supervision model 
is designed to aim
to reconstruct data from masking. 
Then, an RUL prediction model will be built based on the masked self-supervision model. 
For both the masked self-supervision model and the resulting RUL prediction model, 
a transformer architecture is adapted and invoked. 
The overall framework is shown in
Fig.\ref{fig1:Architecture of Masked Self-Supervision for RUL Prediction}.
The effectiveness and applicability of 
the methodology developed here is verified via
our experiments and results described in the manuscript,
and these 
are done on the NASA Commercial Modular Aero-Propulsion System Simulation (C-MAPSS) dataset \cite{cmapss}.


The main contributions of this work are: 
\begin{itemize}
    \item We propose (and fully develop) a method for RUL prediction based on masked self-supervision which can improve the data efficiency, and also extend towards a possible wider applicability. 
    \item Via experiments and results on the NASA Commercial Modular Aero-Propulsion System Simulation (C-MAPSS) dataset \cite{cmapss}, we demonstrate that this methodology utilizing self-supervision can have a significant improvement to the performance for RUL prediction.
\end{itemize}

The overall structure of the remaining paper is organized as follows: In section \ref{sec:proposed_methodology}, we describe the proposed model structure. After that, we present the experiments details, results and analysis in section \ref{sec:experiments}. Finally, section \ref{sec:conclusion} concludes the work with pertinent summary observations.

\section{Proposed Methodology \label{sec:proposed_methodology}}
\subsection{Problem Statement}
The standard RUL prediction model can only be trained with labeled data. The scarcity of broken or faulty machines causes insufficiency in labeled data. The supervised model trained with not enough data could have a risk of overfitting and underrepresented feature extraction. Besides, a great number of unlabeled data which can be collected from healthy machines whenever we want have not been utilized. Regarding the data imbalance between labeled and unlabeled data, we proposed a masked self-supervision method to enhance the RUL prediction with the utilization of the unlabeled data. The network architecture is shown in Fig.\ref{fig:Encoder and Decoder}. 

The RUL prediction problem is formulated as a two-phase learning task in this work. In the first masked self-supervision phase, a masked autoencoder model $F_{p}(\mathcal{X}_{ul}|\mathbf{W_{enc}}, \mathbf{W_{dec}}, \mathbf{\Theta})$ is pretrained on unlabeled data where $\mathcal{X}_{ul}$ is the unlabeled input set, $\mathbf{W_{enc}}$ denotes the encoder weights,  $\mathbf{W_{dec}}$ denotes the decoder weight and $\mathbf{\Theta}$ denotes all the other parameters in the autoencoder model. In the second RUL prediction phase, a prediction model $F_{t}(\mathcal{X}_{l}, \mathbf{W_{enc}}|\mathbf{W}, \mathbf{\Theta})$ is trained on labeled data where $\mathcal{X}_{l}$ is the input set with annotations, $\mathbf{W}$ denotes the weights which is updated based on the pretrained encoder weights $\mathbf{W_{enc}}$.

\subsection{Masked Self-Supervision}
The pretrain model is a masked autoencoder (MAE). Similar with all the other autoencoders, it consists with an encoder and a decoder. The encoder is to map the observed information to a latent representation. After that, decoder would reconstructs the original input from the latent representation. 



We use $\mathbf{X}_{ul} \in \mathbb{R}^{P\times J}$ to denote one sample in the unlabeled input set $\mathcal{X}_{ul}$ where $P$ is the number of timestamps in each sample and $J$ is the dimension of features. $\mathbf{X}_{ul}$ would be divided in to $N$ patches. $\mathbf{H_{ul}^n}\in\mathbb{R}^{K\times J}$ denotes the $n_{th}$ patch, where $K$ is the number of timestamps inside each patch. This process can be illustrated as $\mathbf{X}_{ul} \leftarrow \mathcal{H}_{ul}=(\mathbf{H}_{ul}^1, \mathbf{H}_{ul}^2,...,\mathbf{H}_{ul}^N)$ where $\mathcal{H}_{ul}$ is the set of all the patches. Then, $M$ patches would be randomly masked. $\mathcal{H}_{ul}^{m}$ is the set of masked patches and $\mathcal{H}_{ul}^{um} $ is the set of unmasked patches. In addition, $\mathcal{H}_{ul} = \mathcal{H}_{ul}^{um} \cup \mathcal{H}_{ul}^{m}$. As the masking list is generated randomly, the inputs would be highly sparse which would also create an opportunity for designing an efficient encoder.


\subsubsection{\textbf{Encoder and Decoder}}
Our encoder and decoder used the same model architecture which is a Transformer with a gated convolutional unit. Transformer can capture both long and short-term dependencies to reconstruct the multi-sensor data. In addition, \cite{Transformer-gated-conv} has proved that adding the gated convolutional unit which would extract the local feature from raw sensor data and map it to distributed semantic representations could give a better performance on RUL prediction problem compared with directly applying transformer. After getting the semantic representations, the positional embedding could be added based on their positions before masking. Here we use sine and cosine functions with different frequencies to calculate position embedding:
\begin{equation}
\begin{aligned}
p_{i}^{(2 j)} &=\sin \left(i / 10000^{2 j / J}\right) \\
p_{i}^{(2 j+1)} &=\cos \left(i / 10000^{2 j / J}\right)
\end{aligned}
\label{eq:position embedding}
\end{equation}

Where $p_{i}^j$ is denoted as the positional embedding of $j$th feature at $i$th timestamp. $J$ is the dimension of features (The index $j$ runs from 1 to $J$). 

After adding the positional embedding, it would be input to the transformer. The output of encoder has the same dimension as the input to the encoder. However, they are just a partial of patches among the original input data. The input to decoder would be full set of patches which consists with encoded unmasked patches and masked patches. The masked patches would be regarded as $zero$ and appended to the latent representation. The positions of patches are remaining the same as the original pattern. Then, the positional embeddings (Eq.\ref{eq:position embedding}) would be added to all patches in this full set so that they can have their location information in the original input data. After the unmasked patches set sent to the encoder, we can get the latent representation matrix space set. Let $\mathcal{R}$ denote the latent representation matrix space set after encoder process. The encoder process can be denoted as a function $Enc$.
\begin{equation}
    \mathcal{R}^{um} = Enc(\mathcal{H}_{ul}^{um})\\
\end{equation}

Then, the masked token would be restored to the latent representation matrix. However, the value of masked token would be set to zero or set with interpolation strategy. The process to restore to original dimension can be denoted as: $Res$. 
\begin{equation}
    \mathcal{\hat{R}}^{um \cup m} = Res(\mathcal{R}^{um})\\
\end{equation}

As the dimension of the restored latent representation matrix is the same as original input sample $\mathbf{X}_{ul}$, we can set:
\begin{equation}
    \mathcal{\hat{R}} \leftarrow \mathcal{\hat{R}}^{um \cup m}
\end{equation}

The architecture of decoder is the same as encoder. The decoder can be denoted as $Dec$ and the reconstructed patches set can be denoted as $\mathcal{\hat{H}}$.
\begin{equation}
    \mathcal{\hat{H}}_{ul} = Dec(\hat{R})
\end{equation}
Then, we mapped the reconstructed patches to the input and denoted the reconstructed input as $\mathbf{\hat{X}}_{ul}$.

\begin{equation}
    \mathbf{\hat{X}}_{ul} \leftarrow \mathcal{\hat{H}}_{ul}
\end{equation}

\subsubsection{\textbf{Loss Function}}
For each sample $\mathbf{X}_{ul}$, at $p_{th}$ timestamp, the original multi-sensor data is denoted as $\mathbf{x}_{ul}^p$ and the reconstructed multi-sensor data is denoted as $\mathbf{\hat{x}}_{ul}^p$. There are $P$ timestamp within one sample $\mathbf{X}_{ul}$ (The index of timestamp $p$ runs from $1$ to $P$). In the pretrain model, mean square error (MSE) has been used between the reconstructed and original sample $\mathbf{X}_{ul}$ in the unlabeled input set $\mathcal{X}_{ul}$:

\begin{equation}
\mathcal{L}_{\text {recon }}=\mathbb{E}_{\mathbf{X}_{ul} }\left[\sum_{i=1}^{P}\left\|\mathbf{x}_{ul}^p-\mathbf{\hat{x}}_{ul}^p\right\|_{2}\right]
\end{equation}

\subsection{RUL prediction}
After completing pretraining, the next phase is to do RUL prediction training. The model used for training is still the transformer with a gated convolutional unit. Differently, it would be connected with a regression head in the end to do RUL prediction. In this step, the input data would be labeled input dataset $\mathcal{X}_l$. We denote one sample from $\mathcal{X}_l$ as $\mathcal{X}_l \in\mathbb{R}^{P\times J}$ which has the same dimension as $\mathbf{X}_{ul}$. The patching strategy is the same as the first phase. However, there is no masking conducted. Therefore, the patches form input would be in a full set of patches $\mathcal{H}_l \in\mathbb{R}^{N\times K\times J}$. As we mentioned above, one of the purpose to have pre-training is to give a good initialization for the train model so that it can extract feature better. Therefore, the initial weights for the encoder in train model is inherited from the weights learnt in the encoder of the masked self-supervision model. There are three networks in the RUL prediction model. The first one is a gated convolutional unit for feature extractor which is denoted as ${Con}$. The second one is the transformer which is denoted as $Tr$. The third one is the prediction head which is denoted as $Pr$. We use $g$ to represent the whole networks. Then, for predicting the RUL of a sample $\mathbf{X}_l$ can be written as:

\begin{equation}
\mathbf{\hat{Y}} = g(\mathbf{X}_l) \leftarrow Pr(Tr(Con(\mathbf{X}_l)))   
\end{equation}

Where the $\mathbf{\hat{Y}}$ denotes the predicted RUL and $\mathbf{{Y}}$ denotes the ground truth.

The loss function for phase two is the same as the phase one which is the MSE function. $P$ denotes the total number of timestamps in one input sample $\mathbf{X}_l$(The index of timestamps $p$ runs from 1 to P). Each timestamp is corresponding with one RUL $y$. For each input sample $\mathbf{X}_l$, at $p_{th}$ timestamp, the multi-sensor data is denoted as $\mathbf{x}_l^p$, the ground truth RUL is denoted as $y_l^p$ and the predicted RUL is $\hat{y}_l^p$. Then, the loss function for each input sample $\mathbf{X}_l$ is defined as below:
\begin{equation}
\mathcal{L}_{RUL} =\mathbb{E}_{\mathbf{Y} }\left[\sum_{p=1}^{P}\left\|y_l^{p}-\hat{y}_l^{p}\right\|_{2}\right]
\end{equation}

\section{Experiments \label{sec:experiments}}
There are four sub-datasets in C-MAPSS datasets which are named as FD001, FD002, FD003 and FD004. They are different from the operating conditions and fault modes. In this experiment, only FD001 and FD003 are selected as they have the same operating condition but different fault modes. In future work, we would explore more on the other sub-datasets.

\begin{figure*}[!h]

\subfigure[Masking ratio: 20\%]{
\begin{minipage}[t]{0.23\linewidth}
\includegraphics[width=4.5cm]{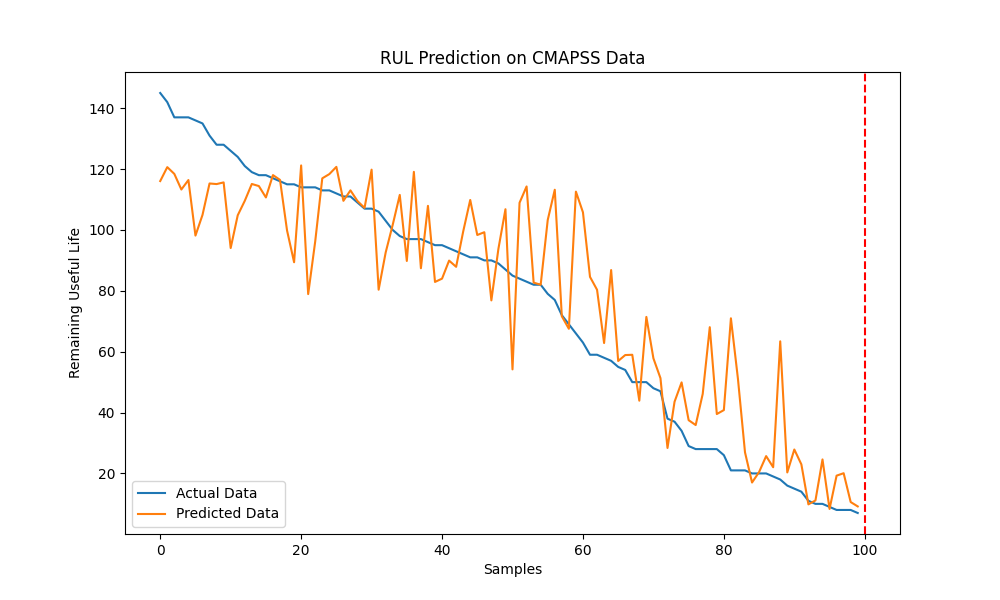}\hfill
\end{minipage}%
}%
\subfigure[Masking ratio: 50\%]{
\begin{minipage}[t]{0.23\linewidth}
\includegraphics[width=4.5cm]{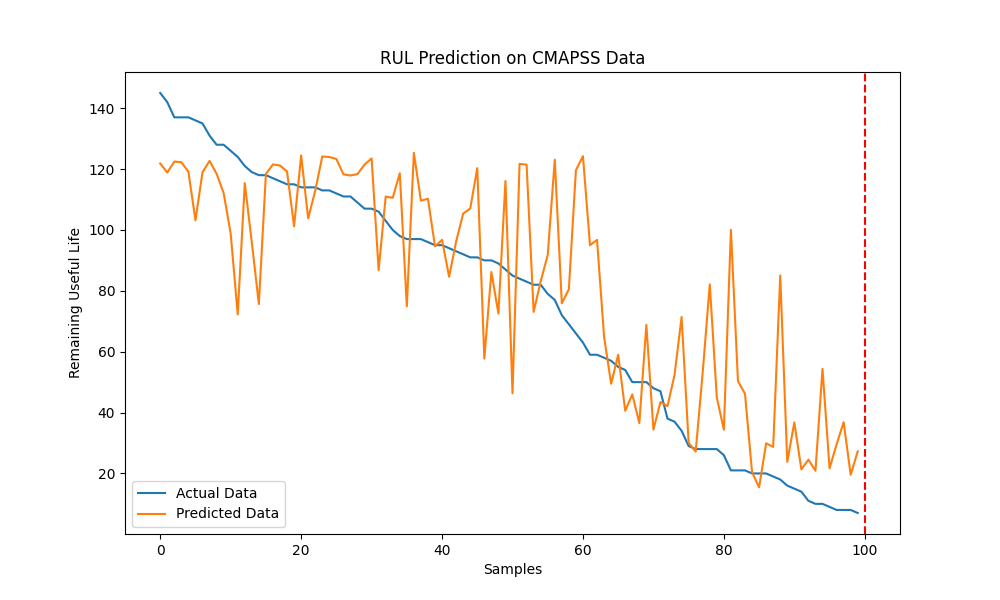}\hfill
\end{minipage}%
}%
\subfigure[Masking ratio: 75\%]{
\begin{minipage}[t]{0.23\linewidth}
\includegraphics[width=4.5cm]{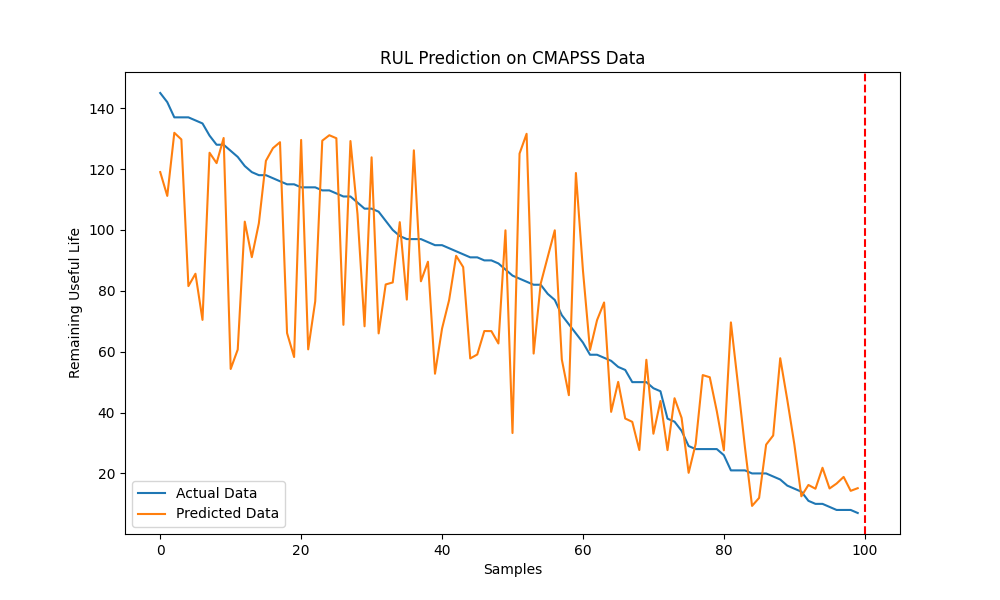}\hfill
\end{minipage}
}%
\subfigure[Without masked self-supervision]{
\begin{minipage}[t]{0.23\linewidth}
\includegraphics[width=4.5cm]{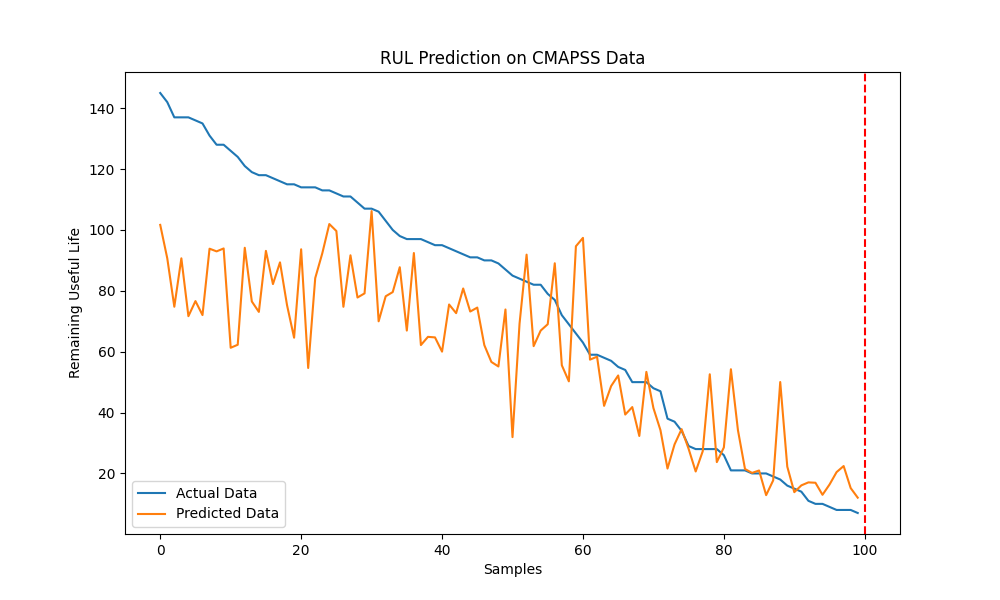}\hfill
\end{minipage}
}%
\caption{Comparison between the actual RULs and the predicted RULs of our model on FD001. All the test units are sorted along the horizontal axis}
\label{fig:comparison FD001}
\end{figure*}

\begin{figure*}[!h]
\subfigure[Masking ratio: 20\%]{
\begin{minipage}[t]{0.23\linewidth}
\includegraphics[width=4.5cm]{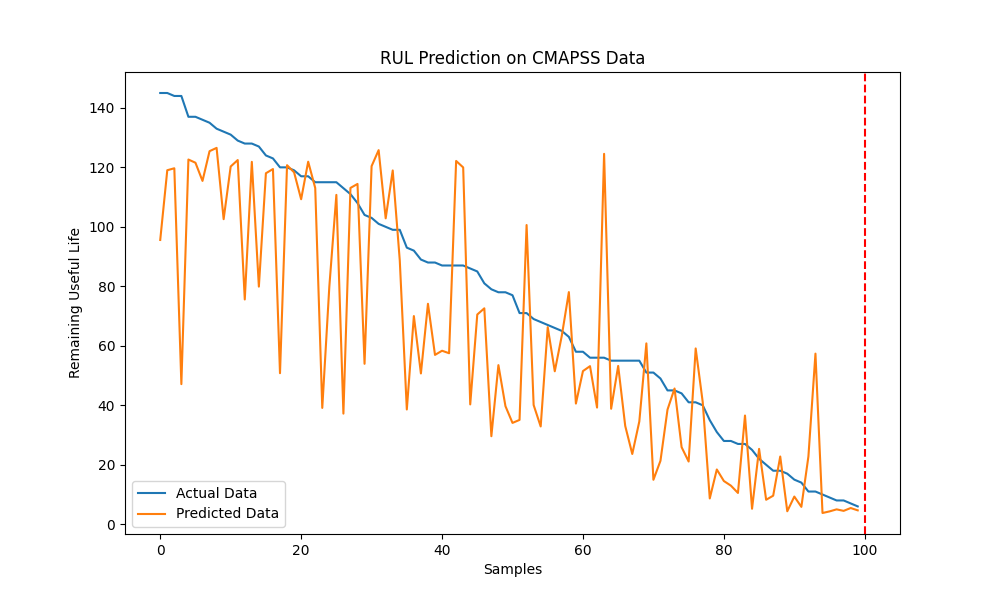}\hfill
\end{minipage}%
}%
\subfigure[Masking ratio: 50\%]{
\begin{minipage}[t]{0.23\linewidth}
\includegraphics[width=4.5cm]{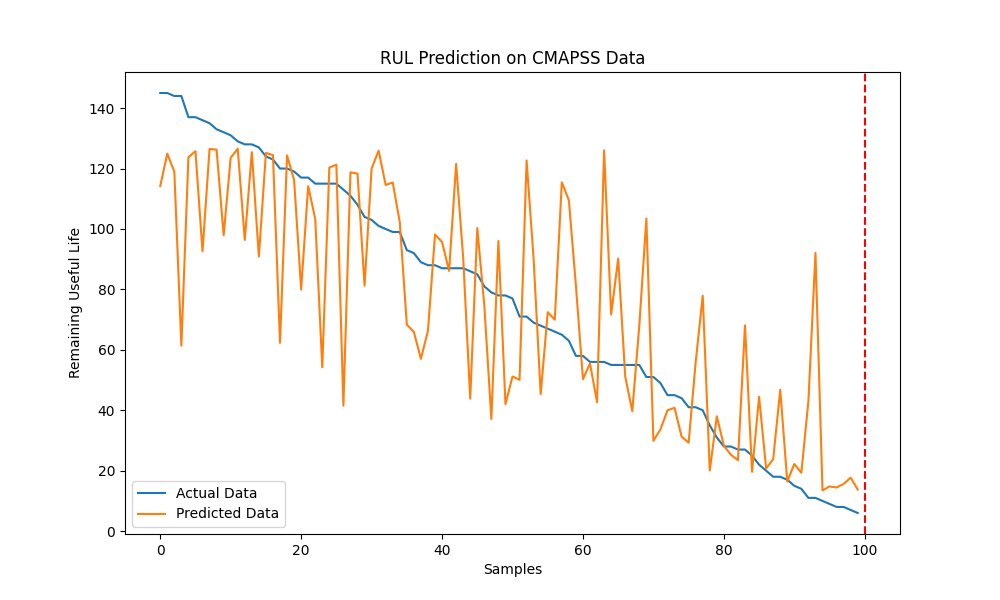}\hfill
\end{minipage}%
}%
\subfigure[Masking ratio: 75\%]{
\begin{minipage}[t]{0.23\linewidth}
\includegraphics[width=4.5cm]{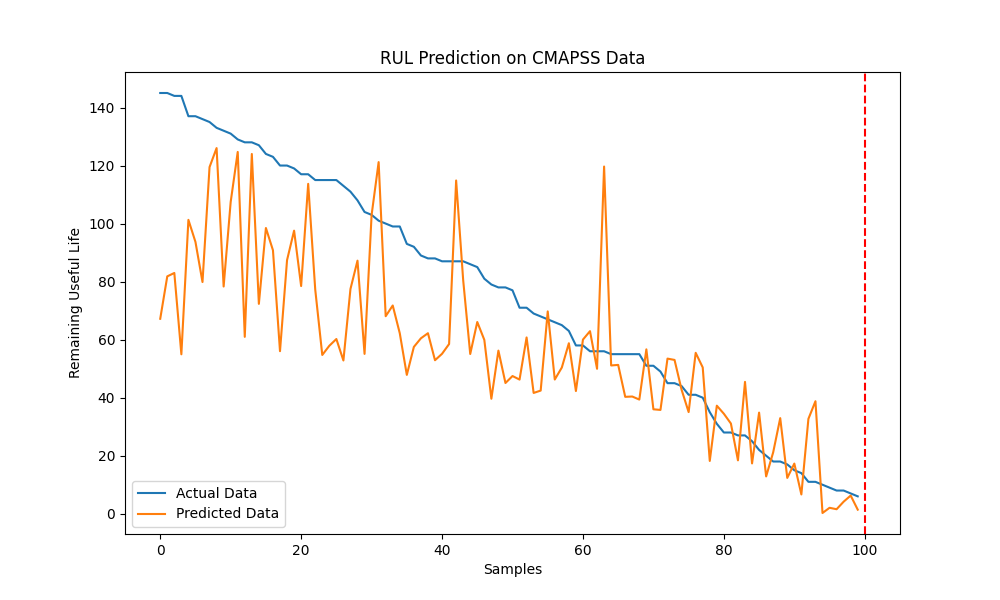}\hfill
\end{minipage}
}%
\subfigure[Without masked self-supervision]{
\begin{minipage}[t]{0.23\linewidth}
\includegraphics[width=4.5cm]{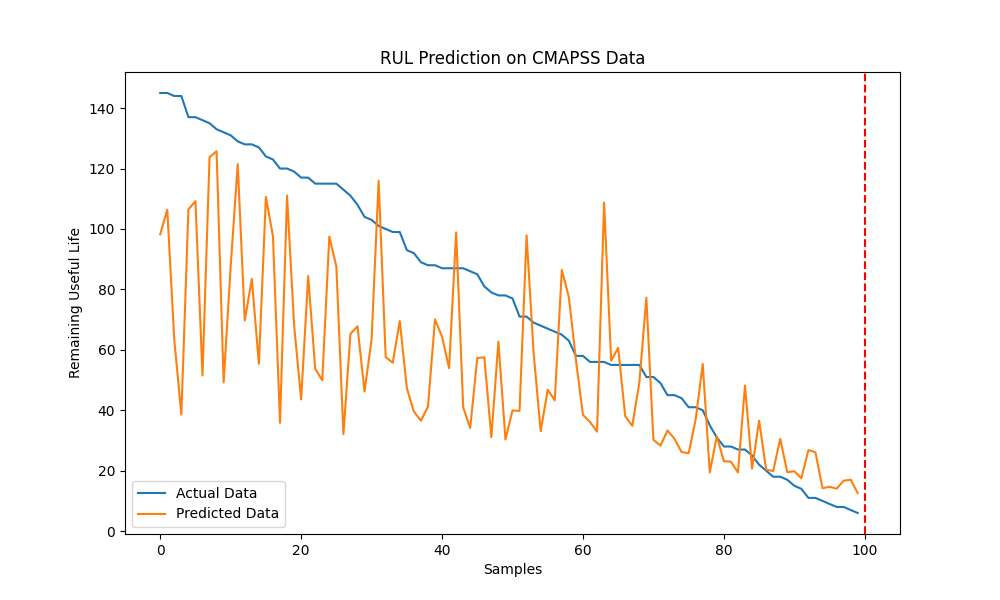}\hfill
\end{minipage}
}%
\caption{Comparison between the actual RULs and the predicted RULs of our model when the masked self-supervision is on FD001 and prediction is on FD003. All the test units are sorted along the horizontal axis}
\label{fig:comparison FD003}
\end{figure*}

\subsection{Data pre-processing}
There are 26 columns included in the dataset. The first two columns are the unit number and operation time. The third column to the fifth column are the operation settings and the other 21 columns are sensor measurements. As not all of the sensor measurements could contribute to the RUL prediction, the first step is to do feature selection. The criteria to reduce the feature dimension is proposed by \cite{data_preprocess_method}. Two metrics are utilized which are the correlation metric and monotonicity metric. The correlation metric describes the linear correlation between features and operation time. The monotonicity metric describes the degeneration information of the features. The correlation function for the $j$th feature is described as:

\begin{equation}
Cor_j =\frac{\left|\sum_{i=1}^{N}\left(f_{i}^{j}-\bar{f}^{j}\right)(i-\bar{i})\right|}{\sqrt{\sum_{i=1}^{N}\left(f_{i}^{j}-\bar{f}^{j}\right)^{2} \sum_{i=1}^{N}(i-\bar{i})^{2}}}
\end{equation}

In addition, the monotonicity metric for the $j$th feature is described as:

\begin{equation}
Mono_j = |\frac{\mathrm{d} f^{j}>0}{N-1}-\frac{\mathrm{d} f^{j}<0}{N-1}|
\end{equation}

Where N is the length of the observation sequence for a whole lifetime. $f^{j}_{i}$ is the $j$th feature at $i$th timestamp. ${\bar{(\cdot)}}$ denotes the mean. Then, the criteria definition for each feature is defined as a linear combination of $Cor_j$ and $Mono_j$\cite{Transformer-gated-conv}:

\begin{equation}
Criteria_j=\gamma \cdot {Cor_j} +(1-\gamma) \cdot Mono_j - \lambda
\end{equation}

$\gamma$ is a tradeoff factor and $\gamma \in[0,1]$. $\lambda$ is a threshold. $Criteria_j$ represents the criteria value for $j$th feature. When $Criteria_j \textgreater 0$, the feature would be chosen. We set $\gamma$ to be 0.5 and $\lambda$ to be 0.2, 14 features out of 21 features are selected.

As the numerical ranges for different sensor measurements are different, after features selection, we did data normalization to normalize the input sensor data within $[0,1]$. The normalization strategy we utilized is min-max normalization:

\begin{equation}
f_{i}^{j} \leftarrow \frac{f_{i}^{j}-\min \mathcal{F}^{j}}{\max \mathcal{F}^{j}-\min \mathcal{F}^{j}}
\end{equation}

The $\mathcal{F}^{j}$ is the whole subset for the $j$th feature. After feature selection and normalization, the next step is to derive the RUL time from operation time. There are $Q$ unit number. The ground truth RUL is derived based on the maximum operation time in each unit number. We denotes operation time for $q$th unit number at $i$th timestamp to be $OP^q_{i}$ and $OP^q_{MAX}$ to be the maximum operation time in $q$th unit number. Then the ground truth RUL for each timestamp in each unit number can be calculated as:

\begin{equation}
RUL^{q}_{i} = OP^q_{MAX} - OP^q_{i}
\end{equation}

\subsection{Data sampling and masking}
The training datasets are grouped with the unit number. Each unit number has inconsistent number of samples. Then, We manually sampled them according to the unit number to be 50 timestamps per input sample. Then, we divided them into patch form and each patch includes 3 timestamps. The $s$th patch would include the timestamps at the positions of $(s-1)$, $s$ and $s+1$. The patching is implemented from the second timestamp and ended at 49th timestamp. The patches are overlapped. We did experiments on the masking ratio of $20\%$, $50\%$ and $75\%$. 

\subsection{Implementation}
Our approach is implemented on PyTorch 1.8.0. The parameters settings for the Transformer with a gated convolutional unit is referring to the settings in \cite{Transformer-gated-conv}. The kernel size of gated convolutional unit is 3 and the padding is 1. For the transformer, the number of encoder and decoder layer is $N = 2$, hidden units $d = 128$ and self-attention heads $H = 4$. The dropout rate is set to be 0.1. The optimizer is Adam \cite{Adam_optim} with learning rate of 0.002 for pretraining and 0.001 for training. 
\subsection{Evaluation metric}
We used the root mean square error (RMSE) to evaluate the performance. Let $L$ be the total number of the test set, $y_i$ to be the $i$th ground-truth RUL and $\hat{y}_i$ to be the $i$th predicted RUL. The evaluation metric is defined as follow:
\begin{equation}
RMSE=\sqrt{\frac{1}{L}\sum_{i=1}^{L}(y_i-\hat{y}_i)^2}    
\end{equation}

\subsection{Results and analysis}
The masked self-supervision models for the two experiments are trained on FD001 and the RUL prediction is trained on both FD001 and FD003. All the other settings are the same for all the experiments

\subsubsection{\textbf{RUL Prediction on FD001}}
Table.\ref{tab:FD001} shows the performance when our proposed method are implemented on the same dataset for both masked self-supervision and RUL prediction and Fig.\ref{fig:comparison FD001} shows the comparison between the actual RULs and the estimated RULs. It shows that if no masked self-supervision implemented, the RMSE would increase by 9.71 from the masked self-supervision model with $20\%$ masking ratio. Besides, we did experiments with different ratio of masking, we found that with the increasing in masking ratio, the performance would be worse but the low masking ratio would have a risk to get overfitting. In addition, even when we increase the masking ratio to be 75\%, the performance for the model with masked self-supervision is still better than the one without masked self-supervision pretrain. Therefore, we prove the feasibility, accuracy and efficiency of our proposed method when the masked self-supervision and RUL prediction implemented on the same dataset.

\begin{table}[H]
\caption{Performance comparison of proposed methods and baseline method (FD001)}
\begin{center}
\begin{tabular}{|c|c|c|c|}
\hline
\textbf{Method}& Masking ratio & RMSE&$\triangle$\\
\hline
 &20\%& \textbf{18.27}& \textbf{+9.71}\\
\cline{2-4}
With masked self-supervision&50\%& 23.92& +4.06\\
\cline{2-4}
&75\% &26.52& +1.46\\
\hline
Without masked self-supervision& -- & 27.98& -- \\
\hline
\end{tabular}
\label{tab:FD001}
\end{center}
\end{table}

\subsubsection{\textbf{RUL Prediction on FD003}}
Different from the first experiment, in the second experiment, although the sub-dataset used for masked self-supervision pretrain does not change, we used another sub-dataset, FD003, to do RUL prediction training. Table.\ref{tab:FD003} shows the evaluation results ($RMSE$) and Fig.\ref{fig:comparison FD003} shows the comparison between the actual RULs and the estimated RULs for test sets. It clearly shows that the models with masked self-supervision outperform the model without masked self-supervision. Furthermore, the model with 50\% masking ratio gives the best performance in this experiment where the $RMSE$ decreases by 9.4 from the baseline model (without masked self-supervision). Compared with the first experiment, the overall improvements are even more significant. In general, masked self-supervision model is efficient in initialize the RUL prediction model with a good feature extractor. It is also feasible when the two phases are conducted on two different sub-dataset. Therefore, we proved the feasibility and efficiency of our proposed method.

\begin{table}[H]
\caption{Performance comparison of proposed methods and baseline method (FD003)}
\begin{center}
\begin{tabular}{|c|c|c|c|}
\hline
\textbf{Method}& Masking ratio & RMSE&$\triangle$\\
\hline
 &20\%&28.44&+8.15\\
\cline{2-4}
With masked self-supervision&50\%&\textbf{27.19}&\textbf{+9.4}\\
\cline{2-4}
&75\% &31.04&+5.55 \\
\hline
Without masked self-supervision& -- &36.59& -- \\
\hline
\end{tabular}
\label{tab:FD003}
\end{center}
\end{table}

\section{Conclusion\label{sec:conclusion}}
In this paper,  
a notable and novel masked self-supervision approach 
is proposed and developed
to do RUL prediction.
The methodology is a strategy 
based on the idea of masked autoencoders
which will utilize unlabeled data to do self-supervision 
to improve the performance of the RUL prediction model. 
In thus the work here,
the
masked self-supervised learning approach 
is developed and utilized;
and this is designed
to seek to build a deep learning model for RUL prediction by utilizing unlabeled data. 
The experiments to verify the effectiveness of this development
are implemented on the C-MAPSS datasets (which is collected from 
the data from the
NASA turbofan engine). 
The results rather clearly show that our development and approach here performs better, 
in both accuracy and effectiveness,
for RUL prediction 
when compared with approaches utilizing
a fully-supervised model.


\section{Acknowledgements}
This research is supported by the National University of Singapore under 
the NUS College of Design and Engineering Industry-focused Ring-Fenced PhD Scholarship programme. 
It is also supported in part by Singapore Institute of Manufacturing Technology (SIMTech), A*STAR.
Additionally,
the authors would like to acknowledge useful discussions with Dr Hendrik Schafstall of
Hexagon, Manufacturing Intelligence Division,
Simufact Engineering GmbH.

\vspace{12pt}

\end{document}